\documentclass[12pt, fleqn]{article}
\usepackage[cp1251]{inputenc}
\usepackage{latexsym,amsfonts,amssymb}
\usepackage{graphicx}
\usepackage{cite}
\usepackage{amsbsy}
\usepackage{amsmath}
\usepackage{epsf}

\newtheorem{remark}{Remark}
\newtheorem{theo}{Theorem}
\newcommand{\bt}{\begin{theo}}
\newcommand{\et}{\end{theo}}
\newcommand{\bd}{\begin{displaymath}}
\newcommand{\ed}{\end{displaymath}}

\newcommand{\be} {\begin{equation}}
\newcommand{\ee} {\end{equation}}
\newcommand{\ba} {\begin{array}}
\newcommand{\ea} {\end{array}}

\newcommand{\p} {\partial}

\newcommand{\vp} {\varphi}

\begin{document}
 \begin{center}
 {\Large \bf Conditional symmetries and exact solutions  \\
 of  nonlinear   reaction-diffusion systems with non-constant diffusivities  }\\
 \medskip

{\bf Roman Cherniha$^{\dag,\ddag}$}
 {\bf and  Vasyl' Davydovych$^\dag$}
 \\
{\it  $^\dag$~Institute of Mathematics, Ukrainian National Academy
of Sciences,\\
 3 Tereshchenkivs'ka Street, Kyiv 01601, Ukraine\\
  $^\ddag$~Department  of  Mathematics,
     National University
     `Kyiv-Mohyla Academy', \\ 2 Skovoroda Street,
     Kyiv  04070 ,  Ukraine
 }\\
 \medskip
 E-mail: cherniha@imath.kiev.ua and davydovych@imath.kiev.ua
\end{center}

\begin{abstract}

 $Q$-conditional symmetries (nonclassical symmetries)
 for the general  class of two-component reaction-diffusion   systems  with non-constant diffusivities  are studied.
 Using the  recently  introduced  notion of  $Q$-conditional symmetries
 of the first type, an  exhausted list of  reaction-diffusion  systems  admitting  such symmetry is derived. The results obtained  for the reaction-diffusion   systems are compared with those for the scalar reaction-diffusion equations. The  symmetries found    for  reducing  reaction-diffusion   systems to two-dimensional dynamical systems, i.e., ODE systems, and finding exact solutions are  applied.
 As result,  multiparameter  families  of exact solutions in the explicit  form
 for a nonlinear   reaction-diffusion  system   with an arbitrary diffusivity are  constructed.
 Finally,   the  application of  the exact solutions
   for solving  a  biologically and physically motivated system
   is presented.



\end{abstract}


\section{\bf Introduction}

The  paper is devoted  to the  investigation of
 the two-component RD systems of the form
 \be\label{1}\ba{l}
 U_t=[D^1(U)U_x]_x+F(U,V),\\
V_t=[D^2(V)V_x]_x+ G(U,V),
 \ea\ee
where
  $U= U(t,x)$ and $V= V(t,x)$ are two  unknown functions representing the  densities
  of populations (cells, chemicals), the pressures in  thin
films, etc.
$F(U,V)$ and $G(U,V)$ are the  given smooth functions describing interaction
between them and environment,
 the functions $D^1(U)$ and $D^2(V)$ are the relevant diffusivities
 (hereafter they are  positive smooth  functions)  and the subscripts $t$ and $x$ denote
differentiation with respect to these variables.
 The class of  RD systems  (\ref{1}) generalizes many well-known nonlinear
second-order models and is used to describe various processes in
physics, biology, chemistry  and ecology (see, e.g., the well-known books
 \cite{ames, mur2, mur2003, okubo}). As a particular case, this system
corresponds to a model for the chemical basis of morphogenesis
proposed by Turing \cite{turing} and is called the interacting
population diffusion system (for two species) \cite [Section
9.2]{mur2}. Usually the diffusivities $D^k \ (k=1,2)$ are taken to
be positive constant, however, in certain insect dispersal models
they depend on the densities $U$ and $V$,
 for example,  a power dependence is  adopted  in  \cite [Section 11.4]{mur2}, \cite{mimura}, and \cite{ch-king4}.

During the last decades  RD systems of the form  (\ref{1})  have been extensively studied by means of different mathematical
methods, including the classical Lie  method.
The search
of Lie symmetries of  the class of RD systems  (\ref{1})   with constant diffusivities, i.e.
 \be\label{1*}\ba{l}
 U_t=d_1U_{xx}+F(U,V),\\
V_t=d_2V_{xx}+ G(U,V)
 \ea\ee  was initiated in the paper
\cite{zulehner-ames}.
 At  present,
one can claim that all possible  Lie symmetries of  (\ref{1}) with
the constant diffusivities
 are completely described
\cite{ch-king,ch-king2}.  In the case of  non-constant diffusivities,
 it has been done in  \cite{ibrag-94}, where  30 RD  systems admitting
 non-trivial Lie algebras of invariance were found.  It turns out that
  many of them are locally equivalent  therefore those 30 systems  can be
  reduced to 10  RD  systems with non-trivial Lie symmetry \cite{ch-king4}.
   In the paper \cite{ch-king06},  this result was extended on RD systems in the
     $(n+1)$ -- dimensional  Euclid space. Note that Lie symmetries of  RD systems
      (\ref{1*})  with the  linear cross-diffusion were described in \cite{niki-05}.

In contrary to the Lie symmetry classification problem,  one of $Q$-conditional
symmetry  classification for  the class of RD systems  (\ref{1})  is not
solved at the present time. To the best of our knowledge the most general
 result was derived in \cite{ch-pli-08}, where the  $Q$-conditional symmetries
  (non-classical symmetries) of  the subclass
 \be\label{1**}\ba{l}
 U_t=(U^kU_x)_x+F(U,V),\\
V_t=(V^lV_x)_x+ G(U,V)
 \ea\ee
 where  described (here $k$ and $l$ are real constants). Note that the result obtained therein is incomplete because the  system of determining equations (16) \cite{ch-pli-08} was solved only under additional restrictions.
 The main reason of  such  incompleteness   follows from the structure of  determining equations, which  are essentially nonlinear in contrary to those in the case of  the Lie symmetry classification problem.
Thus,
  to solve $Q$-conditional symmetry  classification problem
 for  the class of RD systems  (\ref{1}), one should look for  new constructive approaches helping to  solve  the relevant  nonlinear system of determining equations.
 A possible approach was recently proposed in   \cite{ch-2010}  and   is used  in this paper.

The paper is organized as follows.
  In  section 2,   two different  definitions of   $Q$-conditional
   invariance are presented, the system of   determining equations is derived  and the main theorem is proved.
  In section 3, the $Q$-conditional  symmetries obtained for  reducing of    RD systems to systems of  ODEs  are applied  and  multiparameter  families  of exact solutions are constructed. An example of application of the exact solution obtained for solving a boundary value problem with the zero-flux  conditions  is presented, too. Finally, we  summarize and discuss
   the results obtained   in
the last section.

\section{\bf Conditional symmetry for RD systems }

\subsection{Definitions and preliminary analysis}

One notes that   the
RD system  (\ref{1}) can be simplified by applying  the Kirchhoff
substitution \be\label{2} u = \int D^1(U)dU, \quad v = \int
D^1(V)dV, \ee where $u(t,x)$ and $v(t,x)$ are new unknown functions.
Hereafter we assume that there exist unique  inverse functions to those
arising in right-hand-sides of (\ref{2}). Substituting (\ref{2})
into (\ref{1}), one obtains \be\label{3}\ba{l}
u_{xx}=d^1(u)u_t+C^1(u,v),
\\v_{xx}=d^2(v)v_t+C^2(u,v),\ea\ee
where the functions  $d^1,\ d^2$ and $C^1, C^2$ are uniquely defined
via $D^1,\ D^2$ and $F, G$, respectively.
In fact,  the formulae  \be\label{3*}d^1(u)=\frac{1}{D^1(U)}, \
d^2(v)=\frac{1}{D^2(V)}, \ C^1(u,v)=-F(U,V), \ C^2(u,v)=-G(U,V),\ee
where  $U =D^1_*(u) \equiv \Bigl(\int D^1(u)du \Bigl)^{-1} , \quad  V=D^2_*(v) \equiv \Bigl(\int D^2(v)dv \Bigl)^{-1}$ (the upper subscripts $-1$ mean  inverse functions).
Hereafter we  construct conditional symmetries for class of RD systems   (\ref{3}) instead of  (\ref{1}).  Having  any conditional symmetry operator of a RD system of the form  (\ref{3}),
one may easily transform those  into the relevant operator and a  RD system from the class  (\ref{1}) provided the inverse functions in  (\ref{3*}) are known.

Here we use the definition of $Q$-conditional symmetry of the first
type  introduced  recently  in  \cite{ch-2010}  and applied successfully  to  search such symmetries for the classical Lotka-Volterra system   \cite{ch-dav-2011}.

It is well-known that to find Lie invariance  operators, one needs
to consider   system (\ref{3}) as the manifold
${\cal{M}}=\{S_1=0,S_2=0 \}$  where \be\label{2-1}  \ba{l}
 S_1 \equiv \ u_{xx}-d^1(u)u_t-C^1(u,v),\\
S_2 \equiv \ v_{xx}-d^2(v)v_t-C^2(u,v), \ea\ee in the prolonged
space of the  variables: $t, x, u, v, u_t, v_t$,$ u_{x}, v_{x},
u_{xx}, v_{xx}, u_{xt}, v_{xt}, u_{tt}, v_{tt}.$ According to the
definition, system (\ref{3}) is invariant under the transformations
generated by the infinitesimal operator \be\label{2-2} Q = \xi^0 (t,
x, u, v)\p_{t} + \xi^1 (t, x, u, v)\p_{x} +
 \eta^1(t, x, u, v)\p_{u}+\eta^2(t, x, u, v)\p_{v},  \ee
if the following invariance conditions are satisfied: \be\label{2-3}
\ba{l} \mbox{\raisebox{-1.1ex}{$\stackrel{\displaystyle
Q}{\scriptstyle 2}$}} S_1
 \equiv  \mbox{\raisebox{-1.1ex}{$\stackrel{\displaystyle
Q}{\scriptstyle 2}$}}\big ( u_{xx}-d^1(u)u_t-C^1(u,v) \big)
\Big\vert_{\cal{M}}=0, \\[0.3cm]
\mbox{\raisebox{-1.1ex}{$\stackrel{\displaystyle Q}{\scriptstyle
2}$}} S_2
 \equiv  \mbox{\raisebox{-1.1ex}{$\stackrel{\displaystyle
Q}{\scriptstyle 2}$}} \big(v_{xx}-d^2(v)v_t-C^2(u,v) \big)
\Big\vert_{\cal{M}}=0. \ea \ee
The operator $ \mbox{\raisebox{-1.1ex}{$\stackrel{\displaystyle  
Q}{\scriptstyle 2}$}} $  
is the second  
 prolongation of the operator $Q$, i.e.  
\be\label{2-4}  
\mbox{\raisebox{-1.1ex}{$\stackrel{\displaystyle  
Q}{\scriptstyle 2}$}}  
 = Q + \rho_t^1\frac{\partial}{\partial u_{t}}+
 \rho_t^2\frac{\partial}{\partial v_{t}}+  
\rho^1_x\frac{\partial}{\partial u_{x}}+ \rho^2_x\frac{\partial}{\partial v_{x}}  
+\sigma_{xx}^1\frac{\partial}{\partial u_{xx}}  
+\sigma_{xx}^2\frac{\partial}{\partial  v_{xx}},  
\ee  
where the coefficients $\rho$ and $\sigma$ with relevant subscripts  
are expressed  via the functions $\xi^0, \xi^1, \eta^1$ and $\eta^2$
by well-known formulae (see, e.g., \cite{fss, olv, b-k}).

The crucial idea used for introducing the notion of $Q$-conditional
symmetry (non-classical symmetry) is to change  the manifold
${\cal{M}}$, namely: the operator $Q$ is used to reduce ${\cal{M}}$ (see the pioneer  paper  \cite{bl-c}).
However,
 there are two essentially different
possibilities to realize this idea in the case of two-component
systems. Moreover, there are many different possibilities in the
case of multi-component systems  \cite{ch-2010}. Following
\cite{ch-2010}, we formulate two definitions  in the case of system
(\ref{3}).

\noindent \textbf{ Definition 1.} Operator (\ref{2-2}) is called the
$Q$-conditional symmetry of the first type  for the RD system
(\ref{3}) if  the following invariance conditions are satisfied:
\be\label{2-5} \ba{l}
\mbox{\raisebox{-1.1ex}{$\stackrel{\displaystyle Q}{\scriptstyle
2}$}} S_1
 \equiv  \mbox{\raisebox{-1.1ex}{$\stackrel{\displaystyle
Q}{\scriptstyle 2}$}}\big ( u_{xx}-d^1(u)u_t-C^1(u,v) \big)
\Big\vert_{{\cal{M}}_1}=0, \\[0.3cm]
\mbox{\raisebox{-1.1ex}{$\stackrel{\displaystyle Q}{\scriptstyle
2}$}} S_2
 \equiv  \mbox{\raisebox{-1.1ex}{$\stackrel{\displaystyle
Q}{\scriptstyle 2}$}} \big(v_{xx}-d^2(v)v_t-C^2(u,v) \big)
\Big\vert_{{\cal{M}}_1}=0, \ea \ee where the manifold ${\cal{M}}_1$
is either $\{S_1=0, S_2=0, Q(u)=0 \}$ or $\{S_1=0, S_2=0, Q(v)=0 \}$.

\noindent \textbf{Definition 2.} Operator (\ref{2-2}) is called the
$Q$-conditional symmetry of the second  type, i.e., the standard
non-classical symmetry  for the RD system
(\ref{3}) if  the following invariance conditions are satisfied:
\be\label{2-5} \ba{l}
\mbox{\raisebox{-1.1ex}{$\stackrel{\displaystyle Q}{\scriptstyle
2}$}} S_1
 \equiv  \mbox{\raisebox{-1.1ex}{$\stackrel{\displaystyle
Q}{\scriptstyle 2}$}}\big ( u_{xx}-d^1(u)u_t-C^1(u,v) \big)
\Big\vert_{{\cal{M}}_2}=0, \\[0.3cm]
\mbox{\raisebox{-1.1ex}{$\stackrel{\displaystyle Q}{\scriptstyle
2}$}} S_2
 \equiv  \mbox{\raisebox{-1.1ex}{$\stackrel{\displaystyle
Q}{\scriptstyle 2}$}} \big(v_{xx}-d^2(v)v_t-C^2(u,v) \big)
\Big\vert_{{\cal{M}}_2}=0, \ea \ee where the manifold
${\cal{M}}_2=\{S_1=0, S_2=0, Q(u)=0, Q(v)=0 \}$.

\begin{remark} To the  best of our  knowledge, there are
not many paper devoted to search of $Q$-conditional symmetries for
{\it the systems of PDEs} \cite{ch-pli-08, lou-92,bar-2002,ch-se-03,
murata-06, arrigo2010}.
 One may easily check  that Definition 2 was only used in all these  papers.
\end{remark}

It is easily seen that ${\cal{M}}_2 \subset {\cal{M}}_1 \subset
{\cal{M}}$, hence, each Lie symmetry is automatically a
$Q$-conditional symmetry of the first and second  type, while
$Q$-conditional symmetry of the first type is one of the  second
type. From the formal point of view is enough to find all the
$Q$-conditional symmetry of the second type. Having the full list of
$Q$-conditional symmetries of the second  type, one may simply check
which of them is Lie symmetry or/and  $Q$-conditional symmetry of
the first type.

On the other hand,
 to construct   $Q$-conditional symmetries of both types for  a
system of PDEs, one needs to solve new nonlinear system, so called
system of determining equations, which usually is much more
complicated than one for searching Lie symmetries. As follows from
the paper \cite{ch-pli-08}, the complete description of
$Q$-conditional symmetries of the second  type (non-classical
symmetries) for the class of RD systems  (\ref{3}) is very difficult
problem. The corresponding system of determining equations was not
completely solved even for power diffusivity coefficients $d^1(u)$
and $d^2(v)$ (this was done only under additional restrictions on
the form of  non-classical  symmetry operators). It turns out that
the system of determining equations (DEs) to find $Q$-conditional
symmetries of the first   type for  RD systems  of the form
(\ref{3})   is simpler and can be completely  integrated. Having in
the mind  the  {\it complete description} of  $Q$-conditional
symmetries of the first   type, we present here the most interesting
result occurring in the case   $d^1_ud^2_v\neq0$, i.e., both
diffusivity coefficients are arbitrary non-constant functions.

Formally speaking, we should construct  systems of DEs  using  two different manifolds  ${\cal{M}}_1$ (see Definition 1).
However, the class of RD systems  (\ref{3}) is invariant under discrete transformations $u \to v, \, v \to u$. Thus, we can use only the manifold, say,
 $\{S_1=0, S_2=0, Q(u)=0 \}$. Having the complete list of the conditional symmetry operators and the relevant forms of RD systems, one may simply
 extend such list by application the transformations mentioned above.

 Thus, now we present the system of DEs, obtained by direct application of  Definition 1 with
  ${\cal{M}}_1=\{S_1=0, S_2=0, Q(u)=0 \}$:
\begin{equation}\label{4}
 \xi^0_{x}=\xi^0_{u}=\xi^0_{v} =\xi^1_{u}=\xi^1_{v}=0,\end{equation}
\begin{equation}\label{5} \eta^1_{v}=\eta^1_{uu}=\eta^2_{uu}=
\eta^2_{vv}=\eta^2_{uv}=0,\end{equation}
\begin{equation}\label{6} \xi^1\eta^2_u(d^2-d^1)+2\xi^0\eta^2_{xu}=0,\end{equation}
\begin{equation}\label{7}  (\xi_t^0\xi^1-\xi^0\xi^1_t-2\xi^1\xi^1_x)d^1-\xi^1\eta^1d^1_u
-2\xi^0\eta^1_{xu}+\xi^0\xi^1_{xx}=0,\end{equation}
\begin{equation}\label{8} (2\xi^1_x-\xi^0_t)d^2 +\eta^2d^2_v =0,\end{equation}
\begin{equation}\label{9} \xi^1_td^2+2\eta^2_{xv}-\xi^1_{xx}=0,\end{equation}
\begin{equation}\label{10} \frac{\eta^1}{\xi^0} \eta^1d^1_u+(\eta^1_t+2\xi^1_x\frac{\eta^1}{\xi^0}
- \xi^0_t\frac{\eta^1}{\xi^0})d^1 -\eta^1_{xx}+
 \eta^1C^1_u+\eta^2C^1_v+(2\xi^1_x-\eta^1_u)C^1=0,\end{equation}
\begin{equation}\label{11} (\eta^2_t+\frac{\eta^1}{\xi^0}\eta^2_u)d^2-\frac{\eta^1}{\xi^0}\eta^2_u d^1-\eta^2_{xx}
+\eta^1C^2_u+\eta^2C^2_v-\eta^2_uC^1+ (2\xi^1_x-\eta^2_v)C^2=0,
\end{equation}
where  $\xi^0\neq0$. If   $\xi^0=0$ then the system of DEs takes the form
\begin{equation}\label{4*} \xi^1_{u}=\xi^1_{v}=0,\end{equation}
\begin{equation}\label{5*} \eta^1_{v}=\eta^2_{u}=
\eta^2_{vv}=0,\end{equation}
\begin{equation}\label{6*} 2\xi^1_xd^1 +\eta^1d^1_u =0,\end{equation}
\begin{equation}\label{7*} 2\xi^1_xd^2 +\eta^2d^2_v =0,\end{equation}
\begin{equation}\label{8*} \xi^1_td^2+2\eta^2_{xv}-\xi^1_{xx}=0,\end{equation}
\begin{equation}\label{9*}\eta^1_td^1
-\Big(\frac{\eta^1}{\xi^1}\Big)^2\eta^1_{uu}-\frac{\eta^1}{\xi^1}\big(\xi^1_td^1+2\eta^1_{xu}
-\xi^1_{xx}\big)  -\eta^1_{xx}+
 \eta^1C^1_u+\eta^2C^1_v+(2\xi^1_x-\eta^1_u)C^1=0,\end{equation}
\begin{equation}\label{10*} \eta^2_td^2-\eta^2_{xx}
+\eta^1C^2_u+\eta^2C^2_v+(2\xi^1_x-\eta^2_v)C^2=0.
\end{equation}

One notes that  systems of DEs  (\ref{4})--(\ref{11}) ($\xi^0\neq0$)
and (\ref{4*})--(\ref{10*}) ($\xi^0=0$)  are essentially  different
and must be solved separately. Here we restrict ourselves on the
case $\xi^0\neq0$, which is  more complicated.

It should be stressed that we find purely conditional symmetry operators, i.e., exclude all such operators,
which are equivalent to these presented in
  \cite{ch-king4}.  Having  this aim,  we use the system DEs for search Lie symmetry operators,
\begin{equation}\label{12}
 \xi^0_{x}=\xi^0_{u}=\xi^0_{v} =\xi^1_{u}=\xi^1_{v}=0,\end{equation}
\begin{equation}\label{13} \eta^1_{v}=\eta^2_{u}=\eta^1_{uu}=
\eta^2_{vv}=0,\end{equation}
\begin{equation}\label{14} (2\xi^1_x-\xi^0_t)d^1 +\eta^1d^1_u =0,\end{equation}
\begin{equation}\label{15} (2\xi^1_x-\xi^0_t)d^2 +\eta^2d^2_v =0,\end{equation}
\begin{equation}\label{16} \xi^1_td^1+2\eta^1_{xu}-\xi^1_{xx}=0,\end{equation}
\begin{equation}\label{17} \xi^1_td^2+2\eta^2_{xv}-\xi^1_{xx}=0,\end{equation}
\begin{equation}\label{18} \eta^1_td^1 -\eta^1_{xx}+
 \eta^1C^1_u+\eta^2C^1_v+(2\xi^1_x-\eta^1_u)C^1=0,\end{equation}
\begin{equation}\label{19} \eta^2_td^2-\eta^2_{xx}
+\eta^1C^2_u+\eta^2C^2_v+ (2\xi^1_x-\eta^2_v)C^2=0,
\end{equation}
which can be easily derived using the paper   \cite{ch-king4} and substitution  (\ref{2}). One notes, that  systems of DEs
  (\ref{4})--(\ref{11})   and  (\ref{12})--(\ref{19}) coincide if the restrictions
 \be\label{2-5*}
 \eta^2_{u}=0, \quad
(2\xi^1_x-\xi^0_t)d^1 +\eta^1d^1_u =0 \ee take place. Thus, we take
into account only such solutions of  (\ref{4})--(\ref{11}) , which
don't satisfy one of the equations from  (\ref{2-5*}). Moreover,
since $Q$-conditional symmetry of the first type is automatically
one of the  second type, we should also check  the same for
coefficients of the operator obtained by multiplying  (\ref{2-2}) on
any  smooth functions.  Otherwise the  $Q$-conditional symmetry
obtained will be  equivalent to a Lie symmetry.


Now we need to solve the nonlinear system (\ref{4})-(\ref{11}).
Obviously equations  (\ref{4}) and   (\ref{5}) can be easily integrated:\\
 \be\label{2-6}  \xi^0=\xi^0(t), \ \xi^1=\xi^1(t,x), \ \eta^1=r^1(t,x)u+p^1(t,x), \
\eta^2=q(t,x)u+r^2(t,x)v+p^2(t,x), \ee
where $\xi^0(t), \ \xi^1(t,x), \ q(t,x), \ r^k(t,x), \ p^k(t,x) $  ($k=1, 2$) are  arbitrary (at the moment) functions.  Substituting  (\ref{2-6})  into (\ref{8})  and differentiating with respect to
 $u$,   one arrives at  the restriction  $qd^2_v=0$
hence $q=0$  in  (\ref{2-6}).

 Since  $d^2_v\neq0$,  one obtains   $\xi^1_t=0$ from Eq. (\ref{9}), i.e.  $\xi^1$  is a function on  the variable $x$.  Now we need to exam two essentially different cases $\eta^2=0$ and  $\eta^2\neq0$, which follows from
 Eq.  (\ref{8}). It is done  in the next section.

\subsection{The main theorem}

\bt The  RD system (\ref{3}) with $d^1_ud^2_v\neq0$  is
invariant under $Q$-conditional operator of the first type if and
only if   one  and the  corresponding  operator have the forms listed in table
1. Any other RD system admitting  such  $Q$-conditional operator is reduced to one of those from table 1
 by  either  continues  equivalence transformations
\be\label{22}\ba{l}t\rightarrow C_1t+C_2,\\ x\rightarrow C_3x+C_4, \\
u\rightarrow C_5u+C_6,\\ v\rightarrow C_7v+C_8, \ea\ee   with correctly-specified constants $C_l, l=1,\dots,8$ or by discrete transformations
\be\label{22*}
 u \rightarrow v, \quad  v \rightarrow u. \ee
 \et

{\bf Proof.}  To prove   the theorem  one needs to solve   the nonlinear PDE system (\ref{6}) -- (\ref{11}) with restriction  (\ref{2-6}) and $q=0$.  As  follows from the preliminary analysis, we
should  exam  two cases.

\emph{ Let us assume that } $\eta^2=0$. In this case the subsystem  (\ref{6}) -- (\ref{8})
is reduced to the equation
\be\label{41}\xi^1\eta^1d^1_u+2\xi^0r^1_x=0 \ee
because Eq. (\ref{8}) are equivalent to the requirements
$2\xi^1_x-\xi^0_t=0, \, \xi^1_{xx}=0.$  Since   $\eta^1 \not=0$ in Eq.  (\ref{41})
(otherwise   (\ref{2-5*}) will be fulfilled ) we arrive at two subcases
  a) $\xi^1\neq0 \Leftrightarrow r^1_x\neq0;$ b) $\xi^1=0
\Leftrightarrow r^1_x=0$.

\emph{Subcase  a)} doesn't lead to any conditional symmetry operators. In fact, differentiating   (\ref{11})  with respect to  $x$ gives
$(r^1_xu+p^1_x)C^2_u=0 \Rightarrow C^2_u=0.$  Thus, Eq.
(\ref{11}) takes the form  \ $\xi^1_xC^2=0$.

If  $\xi^1_x\neq0$ then  $C^2=0$. Similarly, differentiating  Eq.  (\ref{10})
with respect to  $v, \ u, \ x$ and taking into account that
 $\xi^1\neq0,  \eta^1_x\neq0$  while  $\xi^1_{xx}=0$, we conclude that  $C^1_v=0.$
So,
we arrive at  non-couple RD systems which are excluded from the consideration.

If  $\xi^1_x=0$ then the  corresponding calculations  lead to the
 requirement $r^1=const$, i.e.,  the contradiction to   $r^1_x \not=0$ is obtained.

Thus, the subcase a)  has been  completely  studied.

\emph{Subcase  b)} is more complicated.
First of all  Eq. (\ref{11}) immediately gives   $\eta^1C^2_u=0
\Rightarrow C^2=g(v)$,  where  $g(v)$  is an arbitrary smooth function.
To solve Eq.  (\ref{10}) \be\label{42}
(r^1u+p^1)C^1_u-r^1C^1=p^1_{xx}- (r^1u+p^1)^2d^1_u-
(r^1_tu+p^1_t)d^1, \ee  one needs to exam two subcases:     $r^1\neq0$  and $r^1=0, \
p^1\neq0$.
If
\emph{ $r^1\neq0$ } then the general solution of  (\ref{42}) has the form
\be\label{43}
C^1=(r^1u+p^1)\Big(f(t,x,v)+\int\frac{p^1_{xx}-(r^1_tu+p^1_t)d^1}{(r^1u+p^1)^2}du-d^1\Big),\ee
where  $f(t,x,v)$  is an arbitrary smooth function at the moment and the restriction   $f_v\neq0$
should take place.
Analyzing  the differential consequences of  (\ref{43}): $C^1_{v,x}\equiv
p^1_xf_v+(r^1u+p^1)f_{vx}=0$  and   $C^1_{v,x,u}\equiv
r^1f_{vx}=0$, we  conclude $p^1_x=0, \ f_x=0$.

The differential consequences of    (\ref{43}) with respect to $v$
and  $t$ leads to the equation \be\label{44}
(r^1_tu+p^1_t)f_v+(r^1u+p^1)f_{vt}=0,\ee

Now  it may be shown that   $p^1_t\neq0$  leads to the relation
 $p^1=\alpha r^1,  \alpha \in  \mathbb{R}$.  So, the   $Q$-conditional symmetry operator
has the form  $Q = \p_{t} + r^1(u+\alpha)\p_{u}.$ However, the equivalence transformation
 $u  + \alpha \rightarrow u$,  makes
$p^1=0$. Thus,  assuming  $p^1_t=0$,  we derive from
(\ref{44})  that  $p^1f_{vt}=0$, i.e. either  $p^1=0$  or
$f_{vt}=0 \Rightarrow r^1_t=0$. If   $p^1=0$ then  formula  (\ref{43}) gives
$C^1=uf(v) - \alpha u d^1(u)$, provided  $  r^1=\alpha$. Thus, the case 9 of  table 1 is derived.
Formula   (\ref{43})  with non-constant  $  r^1(t)$ leads to the differential consequence
\[ r^1_td^1_u+\Big(\frac{r^1_t}{r^1}\Big)_t \ \frac{d^1}{u}=0 \]
to find the function $d^1(u)$. Solving this linear ODE, one easily finds   $d^1=\beta u^\alpha$
and the relevant forms for $  r^1(t)$.
Substituting these  expressions  for
$d^1$  and  $r^1(t)$  into  (\ref{43}),   cases  10  and  11 from table 1 are obtained.

The examination of  subcase b) with $r^1=0$  can be  done in a similar way and
cases  12--  15 from table 1  derived.

Thus, the subcase b)  has been  completely  studied.

\emph{ Let us assume that }  $\eta^2\neq0$. Solving Eq. (\ref{8}) with respect to the
function $d^2(v)$  and using the equivalence transformation $v \rightarrow v+C_8$, one obtains two types of the general solution depending on  the function $r^2$ (see the expression for   $\eta^2$ in   (\ref{2-6})):
\be\label{46}d^2=\alpha_1e^{\alpha_2v}, \
\eta^2=\frac{1}{\alpha_2}(\xi^0_t-2\xi^1_x),\ee
\be\label{45}d^2=\alpha_1v^{\alpha_2}, \
\eta^2=\frac{1}{\alpha_2}(\xi^0_t-2\xi^1_x)v,\ee
 where  $\alpha_1$ and $
\alpha_2$  are arbitrary non-zero constants.
Substituting   (\ref{46})  into Eq. (\ref{9}), one concludes that    the equation obtained is equivalent to the conditions
\be\label{45*}
 r^2_x=0, \quad \xi^1_{xx}=0. \ee
 The same result  yields the substitution of (\ref{45}) into Eq. (\ref{9}), excepting the special power
 $\alpha_2 =-4$. We have examined this  power separately and established that  Lie symmetry operators are obtained presented in case 10 of table 1 \cite{ch-king4} because the diffusivity  power $\alpha_2 =-4$ is equivalent to the conformal power $-4/3$ in the case of the RD system   (\ref{1}).

Thus, to solve the  system  of DEs, we need to integrate the equations
  (\ref{7}), (\ref{10}) and  (\ref{11}).  These equations under restrictions derived above take the forms
\begin{equation}\label{47}
\ \xi^1\big((r^1u+p^1)d^1_u+(2\xi^1_x-\xi^0_t)d^1\big)+
2\xi^0r^1_x=0,\end{equation} \be\label{48}\ba{l}
(r^1u+p^1)C^1_u+\eta^2C^1_v=(r^1-\xi^1_x)C^1+r^1_{xx}u+p^1_{xx}-(r^1_tu+p^1_t)d^1-\\
\medskip \qquad
\frac{r^1u+p^1}{\xi^0}\big((r^1u+p^1)d^1_u+(2\xi^1_x-\xi^0_t)d^1\big),\\
(r^1u+p^1)C^2_u+\eta^2C^2_v=(r^2-\xi^1_x)C^2+p^2_{xx}-(r^2_tv+p^2_t)d^2.\ea\ee
Now one notes that the standard integration procedure leads to three different subcases
 \be\label{49}\ba{l}
1)\ r^1=0,\ p^1\ne0,\\
2)\ r^1=p^1=0,\\
3)\ r^1\ne0 \ea\ee
because derivatives (w.r.t. $u$) of all  unknown functions contain   the term  $r^1u+p^1$ as multiplier.

Consider  the first subcase  $ r^1=0,\ p^1\ne0$. Eq. (\ref{47}) with
$\xi^1\neq0$  gives  the exponential form of the function $d^2$ and
the relevant calculations leads only to  $Q$-conditional symmetry
operators, which are equivalent to Lie symmetry operators derived in
\cite{ch-king4}. Thus,  inserting   $\xi^1=0$  and expressions for
$d^2, \ \eta^2$  from (\ref{46})  into (\ref{48})  and integrating
the  equations obtained, one construct their general solution
\be\label{50}\ba{l} \medskip
C^1(u,v)=f(\omega)+\frac{p^1_{xx}}{p^1}u-\frac{p^1}{\xi^0}d^1+\big(\frac{\xi^0_t}{\xi^0}-
\frac{p^1_t}{p^1}\big)\int d^1du,\\
C^2(u,v)=g(\omega)-\frac{\alpha_1}{\alpha_2}\frac{\xi^0_{tt}}{\xi^0_t}e^{\alpha_2v},\ea\ee
where  $f(\omega)$ and $g(\omega)$  are arbitrary smooth functions ( generally speaking, they  contain the variables $t$ and $x$ as parameters)  while
$\omega=u-\alpha_2\frac{p^1}{\xi^0_t}v$  ($\xi^0_t \not=0$ otherwise $\eta^2=0$ !).

Since left-hand-sides in  (\ref{50}) don't depend on $t$ and $x$,
the equations \be\label{53*}
 \frac{p^1}{\xi^0_t}=\beta_1, \quad  \frac{\xi^0_{tt}}{\xi^0_t}=\beta_2 \ee
($\beta_1$ and $\beta_2$  are non-zero constants)
are immediately obtained if   $g(\omega)$ is not a correctly-specified function.
Solving this ODE system  and making the equivalence transformations,
we arrive at case 3 of table 1.
We use also restriction  $d^1(u)$ ($d^1\neq\lambda e^u$) otherwise both
equations  (\ref{2-5*}) will be fulfilled.

The correctly-specified forms for $g(\omega)$ can be identified from
such  differential consequences of (\ref{50})
\[C^2_{ux}\equiv g_{\omega \omega} \omega_x=0, \quad  C^2_{ut}\equiv g_{\omega \omega}\omega_t=0.\]
If  $g_{\omega \omega}\neq0$  then  again   case 3 of table 1  is obtained.
 If  $g(\omega)$ is a linear function, say,
$g(\omega)=g_1(t,x)+g_2(t,x)\omega$  then  the second equation from
(\ref{50}) immediately gives $g_2(t,x)=0,  g_1=\beta_1$  and
$\frac{\xi^0_{tt}}{\xi^0_t}=\beta_2$.

The similar analysis for the function
 $C^1$  from  (\ref{50}) leads to the requirement  $f(\omega)=f_1(t,x)+f_2(t,x)\omega$,  where
$f_1$ and $ f_2$ are arbitrary smooth functions at the moment, hence
\be\label{50*}C^1=f_1+f_2u-\alpha_2f_2\frac{p^1}{\xi^0_t}v
+\frac{p^1_{xx}}{p^1}u-\frac{p^1}{\xi^0}d^1+\Big(\frac{\xi^0_t}{\xi^0}-
\frac{p^1_t}{p^1}\Big)\int d^1du.\ee
Since  left-hand-side   of (\ref{50*}) cannot depend on   $t$ and $x$, one concludes that
$f_2(t,x)=\beta_3\frac{\xi^0_t}{p^1}$ ($\beta_3$  is a non-zero
constant).
To find the function $d^1(u)$, we used the  differential consequence of (\ref{50*})
\[ C^1_{ut}\equiv
\Big(f_2+\frac{p^1_{xx}}{p^1}\Big)_t +\Big(\frac{\xi^0_t}{\xi^0}-
\frac{p^1_t}{p^1}\Big)_t d^1  - \Big(\frac{p^1}{\xi^0}\Big)_td_u^1
=0,\] which is a linear ODE with respect to $d^1(u)$. It turns out
that only solution of the form $d^1=\beta_4+\beta_5u$ ($\beta_4$ and
$\beta_5\neq0$  are arbitrary constants)  leads to new
 $Q$-conditional symmetry operator. This operator and the corresponding functions  $C^1$ and  $C^2$ are listed in case 16 of table 1.

To complete  the examination of the first subcase, we insert the
expressions for  $d^2$ and $\eta^2$  from (\ref{45})  into
(\ref{48})  and integrate  the  equations obtained. The  general
solution   takes the form
 \be\label{51}\ba{l} \medskip
C^1(u,v)=f(\omega)+\frac{p^1_{xx}}{p^1}u-\frac{p^1}{\xi^0}d^1+\big(\frac{\xi^0_t}{\xi^0}-
\frac{p^1_t}{p^1}\big)\int d^1du,\\
C^2(u,v)=v\big(g(\omega)-\frac{\alpha_1}{\alpha_2}\frac{\xi^0_{tt}}{\xi^0_t}v^{\alpha_2}\big),\ea\ee
where $\omega=\exp\big(-\frac{\xi^0_t}{p^1}u\big)v^{\alpha_2}$.
Assuming that
 $g(\omega)$ is an arbitrary smooth function,
 we again arrive at equations  (\ref{53*})
to find  $\xi^0$ and  $p^1$. Thus, case 4 of table 1 was identified.

Finally, we establishes that the  function $g(\omega)=\beta_1$  only
leads to another   $Q$-conditional symmetry operator. Then the
second equation of   (\ref{51})  ultimately  requires that $
\frac{\xi^0_{tt}}{\xi^0_t}=\beta_2$. Analyzing the differential
consequences  $C^1_{vx}=0, \ C^1_{vt}=0$  for the function  $C^1$
from (\ref{51})  we find
$ f(\omega)=f_1(t,x)+f_2(t,x)\ln(\omega).$

Thus, we arrive at the expression
\be\label{52}C^1=f_1+\alpha_2f_2\ln(v)-f_2\frac{\xi^0_t}{p^1}u
+\frac{p^1_{xx}}{p^1}u-\frac{p^1}{\xi^0}d^1+\Big(\frac{\xi^0_t}{\xi^0}-
\frac{p^1_t}{p^1}\Big)\int d^1du\ee that have a  similar structure
to one from (\ref{50*}), hence , we used the same approach to find the function $d^1(u)$   and find the
 $Q$-conditional symmetry operator listed in case 17 of table 1.

 Thus, the first subcase from  (\ref{49}) is completely examined and cases 3, 4, 16 and  17 are derived. Examination of the second subcase from   Eq. (\ref{49}) is rather trivial because    Eqs. (\ref{47})--(\ref{48})  possess  simple  structures so that cases  cases 5--8  can be easily derived.
 Finally, we have done a detailed study of the the third subcase and found six new  $Q$-conditional symmetry operators and corresponding  RD systems, which are   listed in  cases  1,2, 18--21 of table 1.

The proof is now completed. $\blacksquare$

\begin{remark}  All the  $Q$-conditional symmetries  of the first type
listed in table 1 are  automatically  those  of the  second type,
i.e., non-classical symmetries. Because the operators of
non-classical symmetry are equivalent up to multiplication via
arbitrary smooth function, one may observe that cases 5,6,7, and 8
from table 1 are equivalent to those 10, 13, 11, and 14,
respectively.  It should be stressed that  such multiplication does
not allowed for operators of $Q$-conditional symmetry  of the first
type  \cite{ch-2010}. \end{remark}

\newpage
{\bf Table 1.   $Q$-conditional symmetry operators  of the RD system
(\ref{3})  with \quad$d^1_u(u) \, d_v^2(v)\neq0$.}
The following restrictions are assumed:
 $d^1\neq\lambda u^\alpha$  in cases 1 and 2,   $d^1\neq\lambda
e^{\alpha u}$   in case 3, $d^1\neq\lambda e^u$  in case 4.
\begin{small}
\begin{center}

\begin{tabular}{|c|c|c|c|c|c|
} \hline

  &$d^1(u)$ &  $d^2(v)$  & $C^1(u,v)$& $C^2(u,v)$& $Q$   \\

\hline &&&&&\\
1&$d^1(u)$&$v^\beta$&$u\Big(f(v^\beta
u^{-\alpha})-\frac{1}{\alpha}d^1(u)\Big)$&$v\Big(g(v^\beta
u^{-\alpha})-\frac{1}{\beta}v^\beta\Big)$&
 $e^{t}\big(\partial_t+\frac{1}{\alpha}u\partial_
u+\frac{1}{\beta}v\partial_v\big), \ \alpha\beta\neq0$
\\
\hline

 &&&&& \\ 2&$d^1(u)$&$e^v$&$u\Big(f(e^v
u^{-\alpha})-d^1(u)\Big)$&$g(e^v u^{-\alpha})-\alpha e^v$&
 $e^{\alpha t}\big(\partial_t+u\partial_
u+\alpha \partial_v\big), \ \alpha\neq0$
\\
\hline

 &&&&& \\

3&$d^1(u)$&$e^v$&$f(v-\alpha u)-d^1(u)$&$g(v-\alpha u)-\alpha e^v$&
 $e^{\alpha t}\big(\partial_t+\partial_
u+\alpha \partial_v\big), \ \alpha\neq0$
\\
\hline

 &&&&& \\
4&$d^1(u)$ &$v^{\beta}$ &$f(v^{\beta} e^{-u})-d^1(u)$
&$v\big(g(v^{\beta} e^{-u})-\frac{1}{\beta}v^{\beta}\big)$
&$e^t\big(\partial_t+\partial_ u+\frac{1}{\beta}v\p_v\big), \ \beta\neq0$ \\

\hline
 &&&&& \\

5&$d^1(u)$&$v^{\beta}$&$f(u)$&$v\big(g(u)-\frac{\alpha}{\beta}v^{\beta}\big)$
&$(\lambda+e^{\alpha t})\partial_t+\frac{\alpha}{\beta}e^{\alpha
t}v\partial_v, \ \alpha\beta\neq0$
\\
\hline
 &&&&& \\
6&$d^1(u)$&$e^v$&$f(u)$&$g(u)-\alpha e^v$&$(\lambda+e^{\alpha
t})\partial_t+\alpha e^{\alpha t}\partial_v, \ \alpha\neq0$
\\

\hline
 &&&&& \\
7&$d^1(u)$&$v^{\beta}$&$f(u)$&$vg(u)$
&$t\partial_t+\frac{1}{\beta}v\partial_v, \ \beta\neq0$
\\
\hline
 &&&&& \\
 8&$d^1(u)$&$e^v$&$f(u)$&$g(u)$&$t\partial_t+\partial_v$
\\

\hline
 &&&&& \\
9 &$d^1(u)$&$d^2(v)$&$u\big(f(v)- \alpha
d^1(u)\big)$&$g(v)$&$\partial_t+\alpha u\partial_
u, \ \alpha\neq0$\\
\hline
 &&&&& \\

10&$u^{\beta}$ &
$d^2(v)$&$u\big(f(v)-\frac{\alpha}{\beta}u^{\beta}\big)$&$g(v)$&$\p_t+\frac{e^{\alpha
t}}{\lambda+e^{\alpha t}}\frac{\alpha}{\beta}u\p_u, \
 \lambda\alpha\beta\neq0$\\
 \hline
 &&&&& \\
11&$u^{\beta}$ & $d^2(v)$&$uf(v)$&$g(v)$&$\p_t+\frac{1}{\beta
t}u\p_u, \
\beta\neq0$\\
 \hline
 &&&&& \\
 12 &$d^1(u)$&$d^2(v)$&$f(v)- \alpha d^1(u)$&$g(v)$&$\partial_t+\alpha
\partial_
u, \ \alpha\neq0$\\
\hline
 &&&&& \\
 13&$e^u$ &
$d^2(v)$&$f(v)-\alpha e^u$&$g(v)$&$\p_t+\frac{\alpha e^{\alpha
t}}{\lambda+e^{\alpha t}}\p_u, \
 \lambda\alpha\neq0$\\
 \hline
 &&&&& \\
14&$e^u$ & $d^2(v)$&$f(v)$&$g(v)$&$\p_t+\frac{1}{t}\p_u$\\
 \hline
 &&&&& \\
15&$u$ & $d^2(v)$&$f(v)+\alpha u
$&$g(v)$&$\partial_t+p(x)\partial_u, p''=p^2+\alpha p $
\\
\hline 16&$u$ & $e^v$&$\alpha_1v+\alpha_2u+\alpha_3
$&$\alpha_4-e^v$&$e^{t}(\partial_t+p(x)\partial_u+\p_v), \
\alpha_1\neq0, $
\\ &&&&& $p''=p^2+\alpha_2p+\alpha_1$\\
\hline &&&&& \\ 17&$u$ & $v^{\beta}$&$\alpha_1\ln
v+\alpha_2u+\alpha_3 $&$v(\alpha_4-v^{\beta})$&$e^{\beta
t}(\partial_t+p(x)\partial_u+v\p_v), $
\\ &&&&& $p''=p^2+\alpha_2p+\alpha_1, \ \alpha_1\beta\neq0$\\
\hline
\end{tabular}
\end{center}
\end{small}

\begin{small}
\begin{center}
\begin{tabular}{|c|c|c|c|c|c|
} \hline  18&$u$ &
 $v^\beta$&$\alpha_2v^{\frac{1}{\alpha_1}}+\alpha_3u+\alpha_4-u^2
$&$v(\alpha_5-\alpha_1v^\beta),$&$\exp(\alpha_1\beta
t)\Big(\partial_t+ (p(x)+u)\partial_u+$
\\ &&&&$\alpha_1\beta\neq0$& $\alpha_1v\p_v\Big),  \ p''=p^2+\alpha_3p-\alpha_4$\\
\hline &&&&& \\
19&$u$ &
 $ \ e^v \ $&$ \ \alpha_2e^{\frac{v}{\alpha_1}}+\alpha_3u+\alpha_4-u^2 \
$&$ \ \alpha_5-\alpha_1e^v, \ $&$e^{\alpha_1t}\big(\partial_t+
(p(x)+u)\partial_u+\alpha_1\p_v\big), \  $
\\ &&&&$\alpha_1\alpha_2\neq0$& $p''=p^2+\alpha_3p-\alpha_4$\\
\hline  &&&&& \\
20&$u^{-1}$ & $v^\beta$&$\alpha_1v^{-\beta}+\alpha_2u^{-1}
$&$\alpha_3v$&$t\partial_t-(u+\alpha_2t)\partial_u+\frac{1}{\beta}v\p_v,
\ \alpha_1\alpha_2\beta\neq0$
\\
 \hline &&&&& \\
 21&$u^{-1}$ & $e^v$&$\alpha_1e^{-v}+\alpha_2u^{-1}
$&$\alpha_3$&$t\partial_t-(u+\alpha_2t)\partial_u+\p_v, \
\alpha_1\alpha_2\neq0$
\\
 \hline

\end{tabular}
\end{center}
\end{small}

 \section{\bf Reduction nonlinear RD systems to ODE systems \\ and  constructing exact solutions  }

It is well-known that using  any  $Q$-conditional symmetry (non-classical symmetry), one
reduces  the given system of PDEs  to a
system of ODEs  via  the same procedure as for classical Lie
symmetries.  Since  any   $Q$-conditional symmetry of the first type is automatically  one of the  second
type, i.e.,  the  standard $Q$-conditional symmetry, we  apply  this  procedure  for finding exact solutions.

Thus, to construct an ansatz corresponding to the
given operator $Q$, the system of the linear first-order PDEs
\be\label{53}Q(u)=0,\quad  Q(v)=0\ee
 should be solved. Substituting the ansatz obtained
into the RD system  with correctly-specified coefficients, one obtains
the reduced system of ODEs. Since this
procedure is the same for all operators, we consider   in details only the operator and  system arising in case 1 of table 1.
One sees that  PDEs (\ref{53})  for the operator
 \be\label{55}
Q=e^{t}\big(\partial_t+\frac{1}{\alpha}u\partial_
u+\frac{1}{\beta}v\partial_v\big)\ee  takes the form
\be\label{56}u_t=\frac{1}{\alpha}u,\quad  v_t=\frac{1}{\beta}v,\ee
where $x$ should be involved as a parameter because unknown functions depend on two variables.
The general solution of (\ref{56}) is easily constructed, hence, the ansatz
\be\label{57}u=\vp(x)\exp\Big(\frac{t}{\alpha}\Big),\
v=\psi(x)\exp\Big(\frac{t}{\beta}\Big)\ee
is obtained. Here
  $\varphi(x)$ and $
\psi(x)$  are new unknown functions.
To construct the reduced system, we substitute ansatz  (\ref{57})
into  the RD system in question (see table 1)
\be\label{54}\ba{l} u_{xx}=d^1u_t+u\Big(f(v^\beta
u^{-\alpha})-\frac{1}{\alpha}d^1\Big),
\\
v_{xx}=v^\beta v_t+v\Big(g(v^\beta
u^{-\alpha})-\frac{1}{\beta}v^\beta\Big).\ea\ee
It means that we simply calculate the derivatives
$u_t, \ v_t, \ u_{xx}, \ v_{xx},$  and insert them into (\ref{54}).
After the relevant simplifications one arrives at  the ODEs system
\be\label{58}\ba{l} \varphi''=\varphi
f(\varphi^{-\alpha}\psi^\beta),
\\
\psi''=\psi g(\varphi^{-\alpha}\psi^\beta)\ea\ee
(hereafter $\varphi''=\varphi_{xx}, \psi''=\psi_{xx}$)

Ans\"atze and the corresponding reduced systems for other  operators and  systems arising in  table 1
can be constructed in quite similar way. It should be noted that the results obtained in cases 15--19
will be rather cumbersome because the function $p(x)$ arising  therein cannot be expressed in terms of elementary functions.

In table 2, the   ans\"atze and the  reduced systems
are presented for  cases 1--4 of table 1 because those  contain the most general and interesting for application systems of the nonlinear RD equations. The reader may easily extend this table  for other cases listed in table 1.

{\bf Table 2.  Ans\"atze and reduced systems of ODEs corresponding to cases 1--4 of table 1, respectively.}
\begin{small}
\begin{center}

\begin{tabular}{|c|c|c|c|c|c|
} \hline

  &Ans\"atze&Systems of ODEs\\

\hline &&\\
1&$u=\vp(x)\exp\Big(\frac{t}{\alpha}\Big)$&$\varphi''=\varphi
f(\varphi^{-\alpha}\psi^\beta)$\\&$v=\psi(x)\exp\Big(\frac{t}{\beta}\Big)$&$\psi''=\psi
g(\varphi^{-\alpha}\psi^\beta)$ \\ \hline &&\\
2&$u=\vp(x)e^t$&$\varphi''=\varphi
f(\varphi^{-\alpha}e^{\psi})$\\&$v=\psi(x)+\alpha t$
&$\psi''=g(\varphi^{-\alpha}e^{\psi})$
\\ \hline &&\\
3&$u=\vp(x)+t$&$\varphi''=
f(\psi-\alpha\varphi)$\\&$v=\psi(x)+\alpha t$
&$\psi''=g(\psi-\alpha\varphi)$
\\ \hline &&\\
4&$u=\vp(x)+t$&$\varphi''=
f(\psi^{\beta}e^{-\varphi})$\\&$v=\psi(x)\exp\Big(\frac{t}{\beta}\Big)$
&$\psi''=\psi g(\psi^{\beta}e^{-\varphi})$
\\ \hline
\end{tabular}
\end{center}
\end{small}

One sees that the reduced  systems of ODEs are nonlinear and it is quite implausible
that those are integrable for arbitrary smooth functions $f $ and $g $.
However, these systems can be integrated if the functions $f$ and $g$ are correctly specified.
For example, the ODE system arising in case 1, i.e. system  (\ref{58})  takes the form
 \be\label{59}\ba{l}
\varphi''=\lambda_{11}\varphi+\lambda_{12}\psi,
\\
\psi''=\lambda_{21}\varphi+\lambda_{22}\psi,\ea\ee
if one sets
 $\alpha=\beta$ and  the functions $f$ and $ g$ as follows
$f=\lambda_{11}+\lambda_{12}\varphi^{-1}\psi, \
g=\lambda_{22}+\lambda_{21}\varphi\psi^{-1}$,
where  $\lambda_{ij}$ ($i, j=1,2$) are arbitrary constants.
Assuming $\lambda_{12}\neq0$  (otherwise should be $\lambda_{21}\neq0$
and  one will start from the second equation of (\ref{59})) the function $\psi$ can be expressed
from the first  equation  so that the second equation takes the form
 \be\label{60}
\varphi''''-(\lambda_{11}+\lambda_{22})\varphi''+(\lambda_{11}\lambda_{22}-\lambda_{12}\lambda_{21})\varphi=0.\ee

Since  (\ref{60})  is the 4-th order linear ODE its solutions is constructed  using  roots
 of the algebraic equation
\be\label{61}z^4-(\lambda_{11}+\lambda_{22})z^2+(\lambda_{11}\lambda_{22}-\lambda_{12}\lambda_{21})=0.\ee
Obviously, the roots will depend on the values of constants
$\lambda_{ij}$ ($i, j=1,2$) and the accurate  analysis shows that
nine different  forms of the general solutions  occur. They are
listed in table~3. For example, if
$\lambda_{11}\lambda_{22}-\lambda_{12}\lambda_{21}=0$  then the
roots  of (\ref{61}) are
\[z_1=z_2=0, \
z_3=\sqrt{\lambda_{11}+\lambda_{22}}, \
z_4=-\sqrt{\lambda_{11}+\lambda_{22}}. \]

Thus, the general solution of
 (\ref{60})  has the form  \be\label{62} \varphi(x)=C_1\cos(h x)+C_2\sin(h
x)+C_3x+C_4, \ h=\sqrt{-(\lambda_{11}+\lambda_{22})}, \ee if
$\lambda_{11}+\lambda_{22}<0$;
 \be\label{63} \varphi(x)=C_1\exp(h
x)+C_2\exp(-h x)+C_3x+C_4, \ h=\sqrt{\lambda_{11}+\lambda_{22}}, \ee
if  $ \lambda_{11}+\lambda_{22}>0 $  (hereafter   $C_i \  (i=1 \dots
4)$   are arbitrary constants).

Thus, any pair  ($\varphi, \psi$) from table 3 generates the
four-parameter family of solutions  \be\label{65}\ba{l}
u=\varphi(x)\exp\Big(\frac{t}{\alpha}\Big),
\\
v=\psi(x)\exp\big(\frac{t}{\alpha}\big),\ea\ee for non-linear RD
system  \be\label{66}\ba{l} \medskip
u_{xx}=d^1(u)u_t+\lambda_{11}u+\lambda_{12}v-\frac{1}{\alpha}ud^1,
\\
v_{xx}=v^\alpha v_t+\lambda_{21}u+\lambda_{22}v
-\frac{1}{\alpha}v^{\alpha+1},\ea\ee with the corresponding
restrictions on the coefficients $\lambda_{ij}$ ($i, j=1,2$). It
should be noted that the exact solutions obtained are valid for the
RD system (\ref{66}) with arbitrary diffusion coefficient $d^1(u)$.

Finally, we consider an example of possible application of the solutions obtained.

\textbf{Example. }  System (\ref{66}) with  the power diffusivity
$d^1=u^\gamma$ takes the form
 \be\label{69}\ba{l}
 U_t=(U^kU_x)_x-\frac{1}{k+1}\Big(\lambda_{11}U^{k+1}+\lambda_{12}V^{l+1}
 +\frac{l+1}{l}U\Big),\\
V_t=(V^{l}V_x)_x-\frac{1}{l+1}\Big(\lambda_{21}U^{k+1}+
\lambda_{22}V^{l+1}+\frac{l+1}{l}V\Big).
 \ea\ee
  if one applies the substitution ( a particular case of   (\ref{2}))
 \be\label{2*}   u=U^{k+1}, \  v=V^{l+1}, \quad k=-\frac{\gamma}{\gamma+1},  \ l=-\frac{\alpha}{\alpha+1}, \ k\neq-1, \ l\neq-1. \ee

\begin{remark} The RD system  (\ref{69}) is a particular case of
system (34) \cite{ch-pli-08}.
  In fact, setting $\lambda_{1}=\lambda_{3}=0$  in (34) \cite{ch-pli-08} one arrives at  (\ref{69}).
  However, this interesting case was not analyzed in  \cite{ch-pli-08}.
\end{remark}

 Using the notations  $ \lambda^*_{1i} =-\frac{\lambda_{1i}}{k+1}$ and  $ \lambda^*_{2i} =-\frac{\lambda_{2i}}{l+1}\, (i=1,2)$  system  (\ref{69}) can be rewritten as follows
  \be\label{69*}\ba{l}
 U_t=(U^kU_x)_x- \frac{l+1}{l(k+1)}U+ \lambda^*_{11}U^{k+1}+\lambda^*_{12}V^{l+1}
,\\
V_t=(V^{l}V_x)_x-\frac{1}{l}V +\lambda^*_{21}U^{k+1}+
\lambda^*_{22}V^{l+1}.
 \ea\ee

 This system  can be used as a mathematical model for description of  some real processes.
 For example,   (\ref{69*}) with $k=l=1$, i.e.
  \be\label{69**}\ba{l}
 U_t=(UU_x)_x- U+ \lambda^*_{11}U^{2}+\lambda^*_{12}V^{2},\\
V_t=(VV_x)_x-V +\lambda^*_{21}U^{2}+ \lambda^*_{22}V^{2}.
 \ea\ee
   is a system of Lotka-Volterra
type,
 with porous  diffusivities (see, e.g.,
\cite{mur2}), in which the standard terms $\lambda^*_{12}UV$ and
$\lambda^*_{21}UV$ are replaced by the terms $\lambda^*_{12}V^2$ and
$\lambda^*_{21}U^2$, respectively, and  a negative birth-dead rate
is assumed.

System  (\ref{69**}) with $\lambda^*_{11}+\lambda^*_{21}=0$ and
$\lambda^*_{12}+\lambda^*_{22}=0$ can also be regarded as a  model
for the gravity-driven  flow of thin films of viscous fluid through
two networks of pores (in which the fluid pressures are $U(t,x)$ and
$V(t,x)$, the film heights being proportional to the pressures) in a
porous medium \cite{ch-king4}.  The two networks are connected from
one to  other  with some mass transport  presented by the quadratic
terms, while the linear terms represent  the sinks assumed to be
proportional to the pressures.

\newpage
{\bf Table 3. The general solutions of  (\ref{59}).}

\begin{small}
\begin{center}

\begin{tabular}{|c|c|c|c|c|c|
} \hline

 &Exact solutions & Restrictions \\

\hline
1&$\vp=C_1+C_2x+C_3x^2+C_4x^3$&$\lambda_{11}+\lambda_{22}=0$\\&$
\psi=-\frac{1}{\lambda_{12}}\big(\lambda_{11}C_1
-2C_3+(\lambda_{11}C_2-6C_4)x+
\lambda_{11}C_3x^2+\lambda_{11}C_4x^3\big)$&$\lambda_{11}\lambda_{22}-
\lambda_{12}\lambda_{21}=0$ \\ \hline  2&$\varphi(x)=C_1\cos(h
x)+C_2\sin(h x)+C_3x+C_4$&$\lambda_{11}+\lambda_{22}<0$\\&$
\psi(x)=\frac{\lambda_{22}}{\lambda_{12}}\Big(C_1\cos(h x)+C_2\sin(h
x)\Big)-\frac{\lambda_{11}}{\lambda_{12}}(C_3x+C_4)$&$\lambda_{11}\lambda_{22}-
\lambda_{12}\lambda_{21}=0$
\\&&$h=\sqrt{-(\lambda_{11}+\lambda_{22})}$\\ \hline
3&$\varphi(x)=C_1\exp(hx)+C_2\exp(-hx)+C_3x+C_4$&$\lambda_{11}+\lambda_{22}>0$\\&$
\psi(x)=\frac{\lambda_{22}}{\lambda_{12}}\Big(C_1\exp(hx)
+C_2\exp(-hx)\Big)-\frac{\lambda_{11}}{\lambda_{12}}(C_3x+C_4)$&$\lambda_{11}\lambda_{22}-
\lambda_{12}\lambda_{21}=0$
\\&&$h=\sqrt{\lambda_{11}+\lambda_{22}}$\\ \hline
4&$\vp=\exp(hx)\big(C_1+C_2x\big)+\exp(-hx)\big(C_3+C_4x\big)$&$\triangle \equiv (\lambda_{11}-\lambda_{22})^2+4\lambda_{12}\lambda_{21}$ \\
&$\psi=\frac{1}{2\lambda_{12}}
\Big(\exp(hx)\big((\lambda_{22}-\lambda_{11})(C_1+C_2x)+4hC_2\big)+$&$\triangle=0$, $\lambda_{11}+\lambda_{22}>0$\\&
$\exp(-hx)\big((\lambda_{22}-\lambda_{11})(C_3+C_4x)-4hC_4\big)\Big)$
&$h=\sqrt{\frac{\lambda_{11}+\lambda_{22}}{2}}$
\\ \hline
5&$\vp=\cos(h x)\big(C_1+C_2x)+\sin(h x)\big(C_3+C_4x)$&$\triangle=0$\\
&$\psi=\frac{1}{2\lambda_{12}}\Big(\cos(h x)
\big((\lambda_{22}-\lambda_{11})(C_1+C_2x)+ 4hC_4\big)
+$&$\lambda_{11}+\lambda_{22}<0$\\& $\sin(h x)
\big((\lambda_{22}-\lambda_{11})(C_3+C_4x)-4hC_2\big)\Big)$
&$h=\sqrt{-\frac{(\lambda_{11}+\lambda_{22})}{2}}$
\\ \hline
6&$\vp=C_1\exp(h^{-}x)+C_2\exp(-h^{-}x)+C_3\exp(h^{+}x)+C_4\exp(-h^{+}x)$&$\triangle>0$\\
&$\psi=\frac{1}{2\lambda_{12}}(\lambda_{22}-\lambda_{11}-
\sqrt{\triangle})\big(C_1\exp(h^{-}x)+C_2\exp(-h^{-}x)\big)
+$&$\lambda_{11}+\lambda_{22}>\sqrt{\triangle}$\\&
$\frac{1}{2\lambda_{12}}(\lambda_{22}-\lambda_{11}+
\sqrt{\triangle})\big(C_3\exp(h^{+}x)+C_4\exp(-h^{+}x)\big)$
&$h^{\pm}=\sqrt{\frac{\lambda_{11}+\lambda_{22}\pm\sqrt{\triangle}}{2}}$
\\ \hline
7&$\vp=C_1\cos(h_1x)+C_2\sin(h_1x)+C_3\exp(h_2x)+C_4\exp(-h_2x)$&$\triangle>0$\\
&$\psi=\frac{1}{2\lambda_{12}}(\lambda_{22}-\lambda_{11}-
\sqrt{\triangle})\big(C_1\cos(h_1x)+C_2\sin(h_1x)\big)
+$&$(\lambda_{11}+\lambda_{22})^2<\triangle$\\&
$\frac{1}{2\lambda_{12}}(\lambda_{22}-\lambda_{11}+
\sqrt{\triangle})\big(C_3\exp(h_2x)+C_4\exp(-h_2x)\big)$
&$h_1=\sqrt{\frac{\sqrt{\triangle}-\lambda_{11}-\lambda_{22}}{2}}$
\\&&$h_2=\sqrt{\frac{\lambda_{11}+\lambda_{22}+\sqrt{\triangle}}{2}}$\\ \hline
8&$\vp=C_1\cos(h^-x)+C_2\sin(h^-x)+C_3\cos(h^+x)+C_4\sin(h^+x)$&$\triangle>0$\\
&$\psi=\frac{1}{2\lambda_{12}}(\lambda_{22}-\lambda_{11}-
\sqrt{\triangle})\big(C_1\cos(h^-x)+C_2\sin(h^-x)\big)
+$&$\lambda_{11}+\lambda_{22}<-\sqrt{\triangle}$\\&
$\frac{1}{2\lambda_{12}}(\lambda_{22}-\lambda_{11}+
\sqrt{\triangle})\big(C_3\cos(h^+x)+C_4\sin(h^+x)\big)$
&$h^{\pm}=\sqrt{-\frac{\lambda_{11}+\lambda_{22}\pm\sqrt{\triangle}}{2}}$
\\ \hline
9&$\varphi=\exp(h_1x)\big(C_1\cos(h_2x)+C_2\sin(h_2x)\big)+
$&$\triangle<0$\\
&$\exp(-h_1x)\big(C_3\cos(h_2x)+C_4\sin(h_2x)\big)$
&$h_1=\frac{1}{2}\sqrt{\delta+|\lambda_{11}+\lambda_{22}|
}$\\& $\psi=\frac{1}{2\lambda_{12}}\Big[\exp(h_1x)
\Big(\big(C_1(|\lambda_{11}+\lambda_{22}|-2\lambda_{11})+\sqrt{-\triangle}C_2\big)\cos(h_2x)+$
&$h_2=\frac{1}{2}\sqrt{\delta-|\lambda_{11}+\lambda_{22}| }$
\\&$\big(C_2(|\lambda_{11}+\lambda_{22}|-2\lambda_{11})
-\sqrt{-\triangle}C_1\big)\sin(h_2x)\Big)+
$&$\delta=2\sqrt{\lambda_{11}\lambda_{22}-\lambda_{12}\lambda_{21}}$\\
&$\exp(-h_1x)\Big(\big(C_3(|\lambda_{11}+\lambda_{22}|-
2\lambda_{11})-\sqrt{-\triangle}C_4\big)\cos(h_2x)+
$&\\&$\big(C_4(|\lambda_{11}+\lambda_{22}|-2\lambda_{11})
+\sqrt{-\triangle}C_3\big)\sin(h_2x)\Big)\Big]$&\\
\hline
\end{tabular}
\end{center}
\end{small}
\medskip

\medskip
\begin{figure}[t]\label{Fig-1}
\begin{minipage}[t]{8cm}
  \quad  \quad \centerline{\includegraphics[width=9cm]{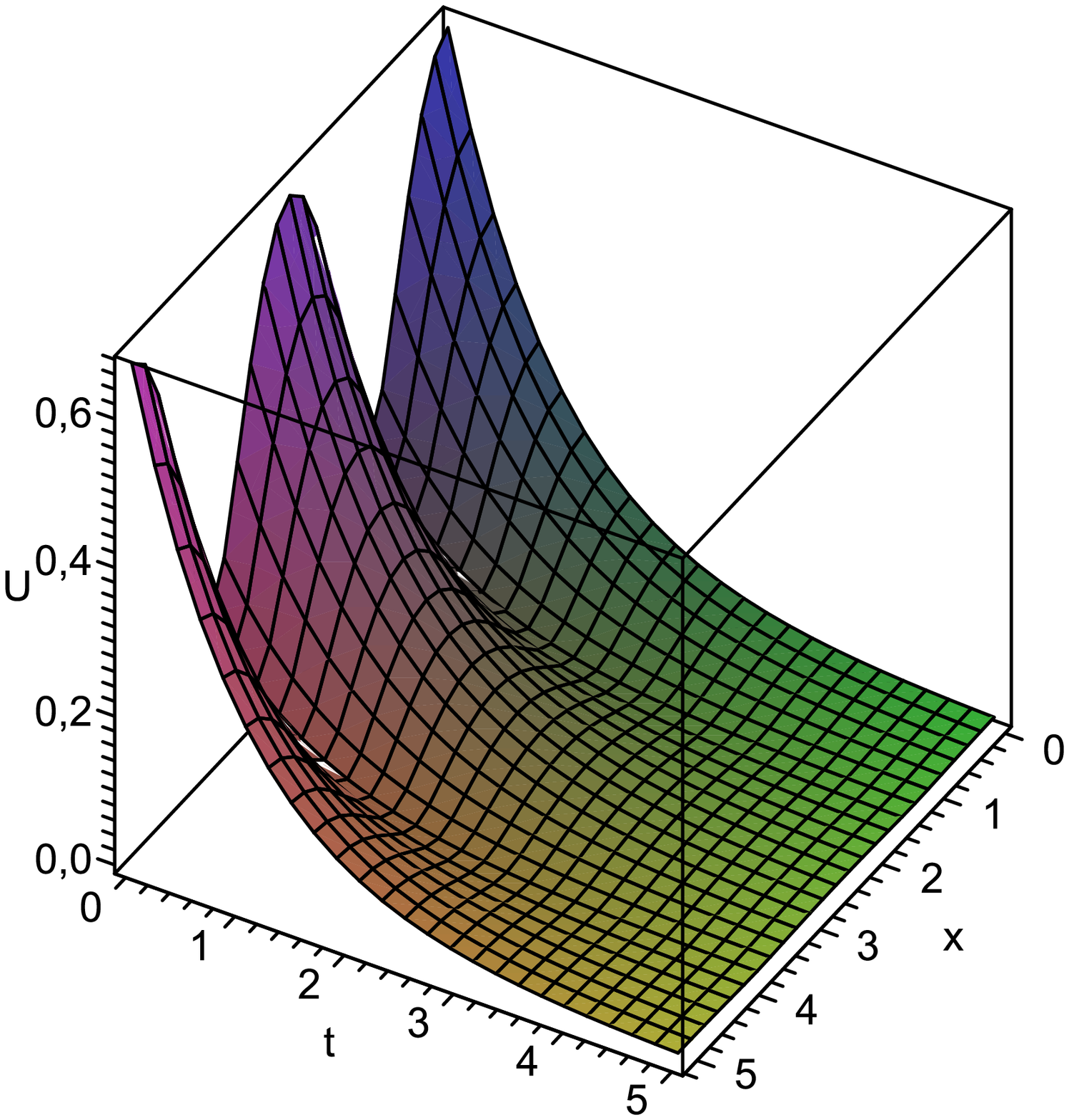}}
\end{minipage}
\hfill
\begin{minipage}[t]{8cm}
\centerline{\includegraphics[width=9cm]{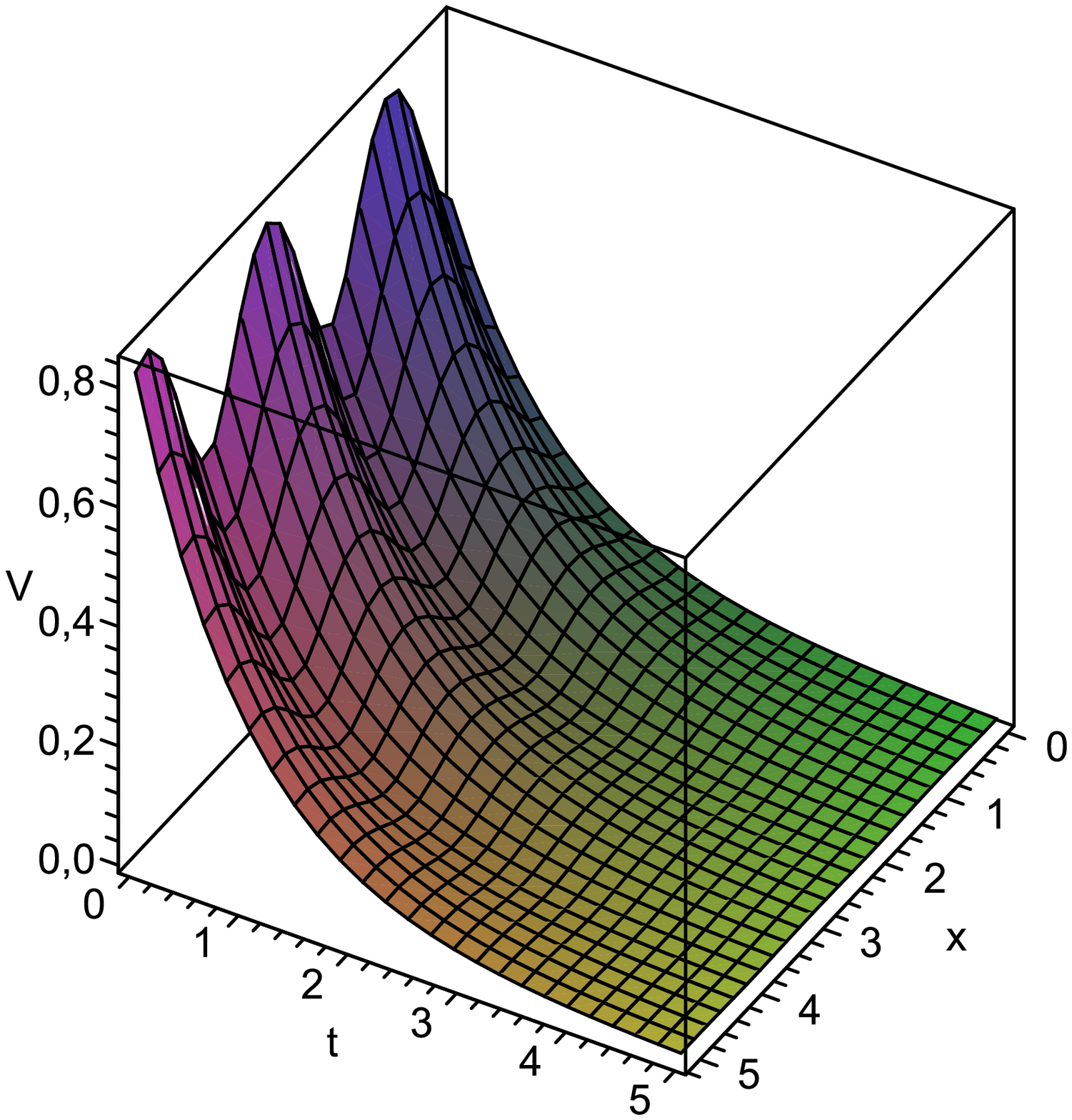}}
\end{minipage}
\center\caption{ Solution  (\ref{70}) 
 with
$ k=l=1, \  \lambda^*_{11}=2, \ \lambda^*_{12}=-1,  \
\lambda^*_{21}=-2, \ \lambda^*_{22}=1, C_1=0.2, \  C_2=0, \
C_4=0.25, \ h=\sqrt{6}.$}
\end{figure}

 Consider the solution listed in case 2 of table 3. Setting   $C_3=0$, one takes the form \be\label{64}\ba{l}
\varphi(x)=C_1\cos(h x)+C_2\sin(h x)+C_4,
\\
\psi(x)=\frac{(l+1)\lambda^*_{22}}{(k+1)\lambda^*_{12}}\Big(C_1\cos(h
x)+C_2\sin(h x)\Big)-\frac{\lambda^*_{11}}{\lambda^*_{12}}C_4,\ea\ee
where $h=\sqrt{(k+1)\lambda^*_{11}+(l+1)\lambda^*_{22}}, \
(k+1)\lambda^*_{11}+(l+1)\lambda^*_{22}>0, \
\lambda^*_{11}\lambda^*_{22}-\lambda^*_{12}\lambda^*_{21}=0.$

Using  the ansatz  (\ref{57}) and substitution  (\ref{2*}),   one obtain the three-parameter family
of solutions
  \be\label{70}\ba{l}
 U=\Big(C_1\cos(h x)+C_2\sin(h
x)+C_4\Big)^{\frac{1}{k+1}}\exp\Big(-\frac{l+1}{l(k+1)}t\Big),\\
V=\Big(\frac{(l+1)\lambda^*_{22}}{(k+1)\lambda^*_{12}}\big(C_1\cos(h
x)+C_2\sin(h
x)\big)-\frac{\lambda^*_{11}}{\lambda^*_{12}}C_4\Big)^{\frac{1}{l+1}}\exp\Big(-\frac{t}{l}\Big).
 \ea\ee

We note that this solution tends to the  stable steady-state point $(0,0)$ of   system (\ref{69*}) provided
 $t\rightarrow\infty$  and the restrictions $l>0, \ k>-1$ take place.
 Moreover,  solution  (\ref{70}) is non-negative and  satisfies the   standard zero-flux conditions on the correctly-specified intervals.
 For instance, solution  (\ref{70}) with $C_2=0$ satisfies the boundary conditions
 \be\label{71}U_x|_{x=0}=0, \ V_x|_{x=0}=0,  \ U_x|_{x=j\frac{\pi}{h}}=0, V_x|_{x=j\frac{\pi}{h}}=0 \ee
  on the space interval $[0, j\frac{\pi}{h}], \  j \in \mathbb{N}$ and its components  are  positive provided
 $\lambda^*_{11}\lambda^*_{12}<0, \ C_4>\max\{|\frac{(l+1)\lambda^*_{22}}{(k+1)\lambda^*_{11}} \, C_1|,  |C_1|\}.$
 Thus, we established that the exact solutions obtained can satisfy the   typical requirements addressed to  physically and  biologically motivated problems.
  For example, the solution  of the   model for the gravity-driven  flow of thin
films  is presented in Fig.1.

\section{\bf Discussion}

In this paper  $Q$-conditional symmetries of
 the class of RD  systems  (\ref {3}) and their application for finding exact solutions
 are studied.
Following the recent paper \cite{ch-2010},    the notion  of   $Q$-conditional symmetry of the first type was used for these purposes.
The main result is presented in Theorem 1 giving an  exhausted list of RD systems of the form   (\ref {3}) with $d^1_ud^2_v\neq0$, which admit such symmetry.
It turns out that there are  exactly 21 RD systems (up to transformations  (\ref{22}) -- (\ref{22*})) admitting $Q$-conditional symmetry  operators of the first type of the form   (\ref{2-2}) with $\xi^0 \not=0$. Note that all the operators found are inequivalent to the Lie symmetry operators presented in \cite{ch-king4}.

It is interesting to compare this result with  the known that  for the single RD equation
 \be\label{1a}
 U_t=\left[D(U)U_x\right]_x + F(U).
\ee
There are several papers devoted to search of  $Q$-conditional (non-classical) symmetry operators  of  equation
 (\ref {1a})  (see \cite {ch-pliu-2006, pop-2010} for details). The  complete  results were derived in   \cite{cla, a-b-h} for
    (\ref {1a}) with constant diffusivity and in   \cite{ a-h} for  (\ref {1a}) with power and exponential diffusivities (note
    that some operators  obtained  in  \cite{ a-h} are equivalent to the Lie symmetry operators). However, there is no complete
    description of  $Q$-conditional symmetry operators with $\xi^0 \not=0$ for  equation  (\ref {1a})
if  $D$ and $F$ are arbitrary smooth function.  In contrary to the single RD equation, we  have done this for the two-component
 RD systems with  arbitrary non-constant diffusivities applying notion of   $Q$-conditional symmetry of the first type.
It turns out that there are 15 systems  of the quite general forms (see cases 1--15 in table 1) admitting such type of symmetry.
On the other hand, there  is no any  single RD equation with the arbitrary   function   $D$ (or  $F$)  admitting $Q$-conditional
(non-classical) symmetry.

The work is in progress to construct all possible symmetries for  the class of RD  system (\ref {3})
with  the  constant diffusivity ($d^1_ud^2_v =0$).
The preliminary analysis  shows that a wide range new RD systems admitting
$Q$-conditional  symmetry operators will be found.

Some  $Q$-conditional operators obtained were used to construct
non-Lie ans\"atze and to reduce the relevant RD systems  to  the
corresponding ODE systems, which are presented in table 2.
Moreover,
 multiparameter  families  of exact solutions in  the explicit  form  (\ref {65})   were  constructed
 for the RD system (\ref {66})  with an arbitrary diffusivity.
 Finally,    application of  the exact solutions
   for solving  the biologically and physically motivated system   (\ref {69*})
   is presented. It  turns out  that the relevant
    boundary value problem at with the zero-flux  conditions
    can be exactly  solved on correctly-specified space intervals.


\end{document}